\documentclass[conference]{IEEEtran}
\IEEEoverridecommandlockouts
\usepackage{cite}
\usepackage{amsmath,amssymb,amsfonts}
\usepackage{algorithmic}
\usepackage{graphicx}
\usepackage{textcomp}
\usepackage{xcolor}
\usepackage{float}
\usepackage{subcaption} %  for subfigures environments 
\def\BibTeX{{\rm B\kern-.05em{\sc i\kern-.025em b}\kern-.08em
    T\kern-.1667em\lower.7ex\hbox{E}\kern-.125emX}}

\usepackage{pifont}
\usepackage{subfig}
\usepackage{colortbl}
\usepackage{braket}
\newcommand{\Gen}{\ensuremath{\mathsf{Gen}}}
\newcommand{\Enc}{\ensuremath{\mathsf{Enc}}}
\newcommand{\Dec}{\ensuremath{\mathsf{Dec}}}
\newcommand{\xmark}{\ding{55}}%
\newcommand{\eloq}{E-LoQ}
\usepackage{balance}

\begin{document}

\title{\eloq: Enhanced Locking for Quantum Circuit IP Protection}
\author{
    \IEEEauthorblockN{Yuntao Liu\IEEEauthorrefmark{1}, Jayden John\IEEEauthorrefmark{2}, Qian Wang\IEEEauthorrefmark{2}}
    \IEEEauthorblockA{\IEEEauthorrefmark{1}Department of Electrical and Computer Engineering, Lehigh University, PA, USA}
    \IEEEauthorblockA{\IEEEauthorrefmark{2}Department of Electrical Engineering, University of California, Merced, CA, USA}
    \IEEEauthorblockA{yule24@lehigh.edu, jjohn92@ucmerced.edu, qianwang@ucmerced.edu}
}

\maketitle

%\title{Quantum Logic Locking (QLL): Safeguarding Intellectual Property in Quantum Circuit Designs}

%\email{webmaster@marysville-ohio.com}
%\affiliation{%
%  \institution{Institute for Clarity in Documentation}
%  \country{USA}
%}

\begin{abstract}
  In recent years, quantum computing has started to demonstrat superior efficiency to classical computing. In quantum computing, quantum circuits that implement specific quantum algorithms are usually not directly executable on quantum computer hardware. Quantum circuit compilers decompose high-level quantum gates into the hardware's native gates and optimize the circuits for accuracy and performance. However, untrusted quantum compilers risk stealing original quantum designs (quantum circuits), leading to the theft of sensitive intellectual property (IP). In classical computing, logic locking is a family of techniques to secure integrated circuit (ICs) designs against reverse engineering and IP piracy. This technique involves inserting a keyed value into the circuit, ensuring the correct output is achieved only with the correct key. To address similar issues in quantum circuit protection, we propose an enhanced locking technique for quantum circuits (E-LoQ) where multiple key bits can be condensed into one key qubit. 
  % to the quantum circuit and use each controlled quantum gate to indicate one key bit of locking key.
  % we propose a method called quantum logic locking, which involves inserting controlled gates to control the function of the quantum circuit. We have 
  Compared to previous work that used one qubit for each key bit, our approach achieves higher security levels. We have demonstrated the practicality of our method through experiments on a set of benchmark quantum circuits. The effectiveness of E-LoQ was measured by assessing the divergence distance from the original circuit. Our results demonstrate that E-LoQ effectively conceals the function of the original quantum circuit, with an average fidelity degradation of less than 1\%.
\end{abstract}

\maketitle

\section{Introduction}
Quantum computing represents a transformative leap in computational power and methodology, with the potential to significantly advance scientific discoveries in fields such as drug discovery \cite{cao2018potential}, material science \cite{bauer2020quantum}, and financial modeling \cite{martin2021toward}. At the heart of quantum computing are qubits, the quantum equivalent of classical bits. Qubits exploit quantum mechanics principles, allowing them to exist simultaneously in multiple states (0, 1, or both), a phenomenon known as superposition. This ability to be in multiple states at once gives quantum computers their extraordinary potential.

In classical computing, logic gates manipulate bits to perform various computations. Similarly, in quantum computing, quantum gates manipulate qubits. Key quantum gates include the Hadamard gate, Pauli gates, controlled gates, and phase gates, which facilitate the execution of complex quantum algorithms. By combining these gates, quantum circuits are formed, enabling computations that connect the software stack (quantum algorithms) to the hardware (quantum processors). Today, multiple cloud providers offer access to quantum technologies, including IBM Quantum \cite{chow2021ibm} Amazon Braket \cite{gonzalez2021cloud}, and Microsoft Azure \cite{prateek2023quantum}. To use a quantum computer, users must submit their designs to a quantum compiler, such as Qiskit \cite{qiskit2024}, which handles tasks like optimization, transpilation, and scheduling of the circuits. Quantum compilers translate high-level quantum algorithms into optimized circuits tailored for specific quantum hardware. However, vulnerabilities exist when malicious compilers or quantum providers could potentially steal quantum designs or engage in unauthorized use of the user's quantum resources. As such, these quantum circuits, composed of multiple quantum gates, are valuable intellectual properties that must be protected from unauthorized use and intellectual property theft.

This paper introduces a method to secure quantum circuits by mimicking traditional logic locking techniques used in electronic circuits. The method involves embedding additional quantum gates within the circuit that require a specific key to function correctly. Without the correct key, the quantum circuit produces incorrect or unpredictable outputs, rendering it non-functional or less effective. The intrinsic entanglement of qubits enables the use of numerous controlled gates within quantum circuits, providing a means to lock or unlock specific functions.
Achieving effective locking for quantum circuits is challenging due to the probabilistic nature of quantum computation.
For instance, if a single-qubit quantum circuit generates the correct output (e.g., 0) in 95\% of measurements and the incorrect output (e.g., 1) in 5\% of measurements, the quantum circuit locking technique must produce a substantially different output distribution.
% This requires the locking technique to 
% address this margin to effectively disrupt functionality. 
% Strategic insertion of random gates is therefore crucial for achieving higher obfuscation.
In addition, the attacker should not be able to recover the original circuit based on the structure of the locked circuit.
% Existing work used a single CNOT gate to represent one key bit to lock circuits like Grover’s algorithm \cite{topaloglu2023quantum}. The lack of variability makes it relatively easy for attackers to guess the correct key and the original circuit.
Existing work used a separate qubit to represent each key bit to lock circuits such as Grover's algorithm \cite{topaloglu2023quantum}. Our approach condenses multiple key bits onto one key qubit, reducing the qubit overhead and allowing higher security levels with a single key qubit.

%demonstrate that quantum circuits can be locked. 
Our proposed scheme incorporates multiple key bits by adding a series of quantum gates at multiple positions in the quantum circuits, thereby expanding the key space. Various gates including Pauli gates, Hadamard gates, and phase gates are employed in the locking process to increase the complexity further for potential attackers. This makes brute-force attacks impractical and ensures that only authorized users with the correct key can access the correct functionality of the quantum circuit, protecting proprietary designs from reverse engineering, overbuilding, and counterfeiting. The main contributions of this paper can be summarized as follows:
\begin{itemize}
   \item We design \textit{\eloq}, a novel locking method for quantum circuits to protect their intellectual property stolen from third-party compilers. Our approach uses a single qubit to incorporate multiple key bits in the locked quantum circuit, which is far more efficient than previous approaches where each qubit can only represent one key bit.
   \item We also propose an unlocking process that restores the correct functionality of the original circuit with the correct key after the circuit has been compiled by the untrusted compiler. The unlocking process removes most of the artifacts added to the circuit in the locking process without impacting the security, making the overhead of \eloq\ close to zero.
   \item We propose a set of metrics for \eloq\ and evaluate \eloq\ using these metrics through extensive experiments on benchmark circuits. Our experiments show that circuits locked with \eloq\ exhibit high functional differences from their original version and are resilient to known attacks.
\end{itemize}

\section{Background \& Related Work}

\subsection{Logic Locking for Classical Circuits}
\label{ssec:logic_locking}
Logic locking is a family of techniques to protect IP in integrated circuits (IC) for classical computing \cite{chakraborty2019keynote}. Since most IC design companies are fabless, they rely on external facilities to manufacture, package, and test the chips. Such an outsourcing process exposes the design details and threatens the security and privacy of the IP. 
Logic locking typically modifies a design by incorporating key inputs and key gates, making the design's functionality dependent on the value of the key input \cite{roy2008epic}. Wrong keys will corrupt the functionality of the design. The designer withholds the correct key during the outsourcing process and only provides it to trusted chip users. This ensures that the chip's true functionality remains hidden from untrusted parties in the IC supply chain.

\begin{figure}[h]
\captionsetup{font=small} 
    \centering
    \includegraphics[width=0.48\textwidth, trim={1mm 7mm 1mm 7mm}, clip]{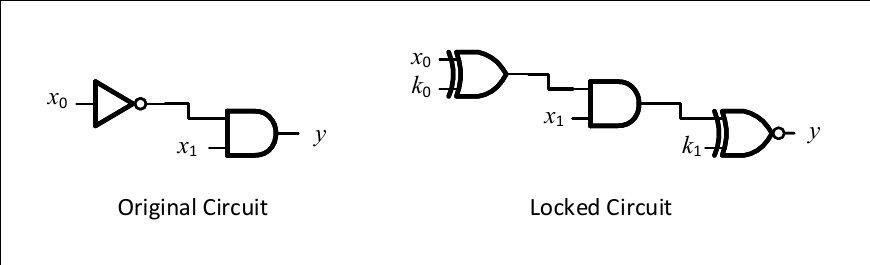}
    \caption{An example of logic locking on classical circuits, On the left, the original circuit is shown with 2-bit input $x_0, x_1$ and 1-bit output $y$; On the right, the locked circuit is displayed, which includes a 2-bit key input $k_0, k_1$ associated with original input $x_0, x_1$; the output $y$ will only be correct if the key ($k_0, k_1$) is correct.}
    \label{fig:locking_example}
\end{figure}

Let us consider the circuits shown in Figure~\ref{fig:locking_example} \cite{liu2021robust}. The original circuit has two input bits $(x_0, x_1)$ and is locked with a two-bit key $(k_0, k_1)$. The output of the locked circuit given each combination of input and key values is shown in Table \ref{tab:locking_truth_table}. Cells with incorrect output values are marked with (\xmark). It can be seen that, given the correct key, all input values will have the correct output. If the key is incorrect, the output will be incorrect for some input values.

\begin{table}[h]
    \centering
    \scriptsize
    \caption{Truth table of the locked circuit in Figure~\ref{fig:locking_example}}
    \begin{tabular}{|c|c|c|c|c|c|}
    \hline
     $(k_0, k_1)$ & $(0,0)$ & $(0,1)$ & \cellcolor{green!25}$(1,1)$ & $(1,0)$ & Correct $y$ \\ \hline
     $\vec{X}=(0,0)$ & 1(\xmark) & 0 & \cellcolor{green!25}0 & 1(\xmark) & 0  \\    
    \hline
     $\vec{X}=(0,1)$ & 1 & 0(\xmark) & \cellcolor{green!25}1 & 0(\xmark) & 1 \\
     \hline
     $\vec{X}=(1,1)$ & 0 & 1(\xmark) & \cellcolor{green!25}0 & 1(\xmark) & 0 \\     
     \hline
     $\vec{X}=(1,0)$ & 1(\xmark) & 0 & \cellcolor{green!25}0 & 1(\xmark) & 0 \\     
     \hline
    \end{tabular}
    \label{tab:locking_truth_table}
\end{table}

Security researchers have investigated numerous attack methods against logic locking. These attacks can be broadly classified into two categories: attacks that require access to a working system and those that do not. For both types of attacks, the attacker is considered to have the netlist of the locked circuit. For attacks that require a working system, it is assumed that the attacker can query an unlocked chip for correct input-output pairs.
The Boolean satisfiability (SAT) based attack \cite{subramanyan2015evaluating} is the most representative attack of this type. Improvements on the original SAT attack include approximate attacks \cite{shamsi2017appsat, shen2017double} and those that incorporate timing \cite{chakraborty2020evaluating} and optical \cite{zuzak2022combined} side-channel information. Counteracting these SAT-based attacks has been a major security goal for most logic locking schemes.
% A set of SAT-resistant obfuscation approaches, such as SARLock~\cite{yasin2016sarlock} and Anti-SAT~\cite{xie2018anti} forced the SAT attacks to undergo exponential complexity. These are also followed by new attacks and newer defenses and the cat-and-mouse game is still ongoing.
Attacks without a working system have to rely on other clues and can usually only narrow down the key search space rather than find the correct key conclusively. For example, the Desynthesis attack \cite{massad2017logic} rules out keys that would not represent an original circuit netlist that can be naturally produced by the synthesis tool. These attacks usually target the structural information leakage of specific logic locking schemes and lack generalizability.

\subsection{Quantum Circuits} \label{ssec:quantum_circuits}
This section provides an overview of the basics of quantum computing, detailing the essential components such as qubits, quantum gates, and quantum circuits. It also emphasizes the critical role of quantum gates, which are central to the implementation of quantum logic locking.

\subsubsection{Qubits}
\begin{figure}[h]
    \captionsetup{font=small} 
    \centering
    \includegraphics[width=0.2\textwidth, trim={1 1 1 1}, clip]{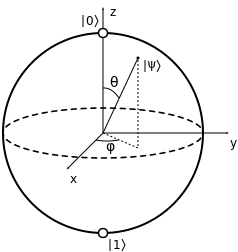}
    \caption{Bloch sphere representation of qubit; Poles represents the classical binary states $\ket 0$ (north pole) and $\ket 1$ (south pole). Any point on the sphere corresponds to a quantum superposition of these states, where the angle on the sphere represents the relative phase and amplitude of the qubit's state.}
    \label{fig:bloch_sphere}
    %\vspace{-0.5cm}
\end{figure}

The quantum bit, or qubit for short, is conceptually similar to the Boolean bits in classical computing. A qubit has two basis states, denoted by the bracket notation as $\ket 0$ and $\ket 1$. Unlike classical Boolean bits, the state of a qubit can be any linear combination of $\ket 0$ and $\ket 1$ with norm 1. Generally, a qubit $\ket \psi$ can be represented as:

\begin{equation*}
  \ket \psi = \alpha \ket 0 + \beta \ket 1,
\end{equation*}
where $\alpha$ and $\beta$ are given by 

\begin{equation*}
  \alpha =\cos \frac{\theta}{2},
  \beta = e^{i\varphi}\sin \frac{\theta}{2},
\end{equation*}

\noindent which are complex numbers satisfying $|\alpha|^2 + |\beta|^2 = 1$. The angles $\theta$ and $\varphi$ can be found in the Bloch sphere representation of a qubit which is shown in Figure \ref{fig:bloch_sphere}.

The basis states for one qubit can be expressed as two-dimensional vectors, e.g., $\ket 0 = [1, 0]^T$ and $\ket 1 = [0, 1]^T$. As a result, the state $\ket \psi$ above can be written as $\ket \psi = \alpha \ket 0 + \beta \ket 1 = [\alpha, \beta]^T$. For multi-qubit states, a similar situation exists. More specifically, there are $2^n$ basis states in the space of $n$-qubit states, ranging from $\ket{0\dots 0}$ to $\ket{1\dots 1}$, and a $n$-qubit state $\ket \psi$ can be expressed by:

\begin{equation*}
  \ket \psi = \sum_{i = 0}^{2^n - 1} a_i \ket i
\end{equation*}
where $\sum_{i = 0}^{2^n - 1}|a_i|^2 = 1$.

\subsubsection{Quantum Gates}

Similar to the operations in classical computing, the operations on qubits in quantum computing are performed by quantum gates. These gates carry out unitary operations, which transform the states of input qubits. Quantum algorithms are composed of sequences of these quantum gates, designed to evolve input qubits into desired quantum states. For example, A quantum gate $U$ must satisfy the equations $U U^\dagger = U^\dagger U = I$, meaning that a quantum gate must be a unitary operation. A quantum gate $U$ operating on a qubit $\ket \psi$ can be written down as $\ket \psi \rightarrow U \ket \psi$. %In the vector-matrix representation, $2^n \times 2^n$ matrices can express $n$-qubit quantum gates. 

\textbf{Single-Qubit Gates:}
Single qubit gates are the fundamental operations in quantum computing, serving as the building blocks for more complex quantum algorithms and circuits. These gates operate on a single qubit, which can exist in a superposition of the classical states 0 and 1. Unlike classical bits, qubits can represent 0, 1, or any quantum superposition of these states, allowing quantum computers to perform parallel computations on an exponential scale.
Single qubit gates are analogous to the elementary logic gates in classical computing, such as the NOT gate. However, their functionality extends far beyond binary flips. These gates manipulate the quantum state of a qubit by rotating its state vector on the Bloch sphere, a geometric representation of the qubit's state. The most common single qubit gates include the Pauli gates (X, Y, Z), the Hadamard gate (H), and phase shift gates (S, T).
\begin{itemize}

    \item \textit{Pauli-X Gate (X):} Equivalent to a classical NOT gate, it flips the state of a qubit from $\ket 0$ to $\ket 1$ and vice versa.
    \item \textit{Pauli-Y Gate (Y):} Combines a bit flip and a phase flip, rotating the qubit state around the Y-axis of the Bloch sphere by  $\pi$.
    \item \textit{Pauli-Z Gate (Z):} Rotating around the Z-axis by $\pi$, effectively leaving the $\ket 0$ state unchanged while flipping the phase of the $\ket 1$ state by $\pi$.
    \item \textit{Hadamard Gate (H):} Creates a superposition of $\ket 0$ and $\ket 1$, transforming a qubit from either base state into an equal probability of being measured as 0 or 1.
    \item \textit{Phase Shift Gates (S, T):} Modifying the relative phase between $\ket 0$ and  $\ket 1$ by $\frac{\pi}{2}$ and $\frac{\pi}{4}$, respectively.
\end{itemize}

% The power of single-qubit gates lies in their ability to precisely control and manipulate qubit states, enabling the construction of more complex quantum operations when combined with multi-qubit gates. Understanding single qubit gates is crucial for exploring the full potential of quantum logic locking, as they serve as the foundation for modifying the functionality of quantum circuits.

%For instance, the Pauli-$X$ gate, a single-qubit gate that flips $\ket 0$ to $\ket 1$ and $\ket 1$ to $\ket 0$, is comparable to the NOT gate in classical computation. One another important example is the CNOT gate, also known as the {\tt CX} gate, which is a two-qubit gate that if the control qubit is in the state $\ket 1$, a Pauli-$X$ gate will be applied to the target qubit, and otherwise nothing will happen. Their matrix representations together with some other matrices of quantum gates are shown below. 

% \begin{equation*}
%   {\tt I}=\begin{bmatrix}
%     1 & 0 \\
%     0 & 1
%   \end{bmatrix},\
%   {\tt X}=\begin{bmatrix}
%     0 & 1 \\
%     1 & 0
%   \end{bmatrix},\
%   {\tt CX} = \begin{bmatrix}
%     1 & 0 & 0 & 0 \\
%     0 & 0 & 0 & 1 \\
%     0 & 0 & 1 & 0 \\
%     0 & 1 & 0 & 0
%   \end{bmatrix},\
%   {\tt RZ}(\theta)=\begin{bmatrix}
%     e^{-i\frac{\theta}{2}} & 0                     \\
%     0                      & e^{i\frac{\theta}{2}}
%   \end{bmatrix},\
%   {\tt SX}=\frac{1}{2}\begin{bmatrix}
%     1+i & 1-i \\
%     1-i & 1+i
%   \end{bmatrix}
% \end{equation*}

\textbf{Multiple-Qubit Gates:} Multi-qubit gates are essential in quantum computing because they enable interactions between qubits, facilitating complex operations that are essential to many quantum algorithms. Unlike single qubit gates, which only affect individual qubits, multi-qubit gates manipulate the states of two or more qubits simultaneously, creating entanglement—a uniquely quantum phenomenon where qubit states become interdependent. This interdependence is critical for harnessing the full power of quantum computation, as it allows for parallel processing and the execution of quantum algorithms that surpass classical methods. Additionally, this interdependence is key for implementing logic locking, as the state of one gate depends on controlled signals generated by other gates.

Gates like the Controlled-NOT (CNOT) are fundamental in constructing these multi-qubit operations, enabling conditional logic where the state of one qubit influences the operation on another. The CNOT gate operates on two qubits: a control qubit and a target qubit. The control qubit would determine whether the operation is applied to the target qubit. The target qubit's state is flipped by the control qubit in the $\ket 1$ state.

Its function is to flip the state of the target qubit (apply an X gate) if, and only if, the control qubit is in the $\ket 1$ state. If the control qubit is in the $\ket 0$ state, the target qubit remains unchanged. this can be expressed in the following truth table:

\begin{table}[h]
\centering
\begin{tabular}{|c|c|c|}
\hline
Control Qubit & Target Qubit (Input) & Target Qubit (Output) \\ \hline
0                      & 0                            & 0                             \\ \hline
0                      & 1                            & 1                             \\ \hline
1                      & 0                            & 1                             \\ \hline
1                      & 1                            & 0                             \\ \hline
\end{tabular}
\caption{Truth Table for the CNOT Gate}
\label{tab:cnot_truth_table}
\end{table}

The CNOT gate is important in quantum computing, particularly in the context of quantum circuit locking, for several reasons. First, it is a frequently used gate to create entanglement. This entanglement is a crucial resource in obfuscating quantum circuits as it allows for conditional operations in quantum circuits which makes it possible to use secret keys to control quantum circuit functionality. 
% Additionally, the CNOT gate, in combination with single qubit gates, forms a universal gate set, meaning that any quantum operation can be decomposed into a sequence of CNOT and single-qubit gates. Furthermore, many quantum algorithms, including Shor's algorithm and Grover's algorithm, rely heavily on the CNOT gate for their implementation, especially in constructing entangled states and performing conditional operations, underscoring its essential role in advancing quantum computation.

\begin{figure}[h]
\centering
\begin{subfigure}{0.5\linewidth}
  \centering
  \includegraphics[width=0.45\linewidth]{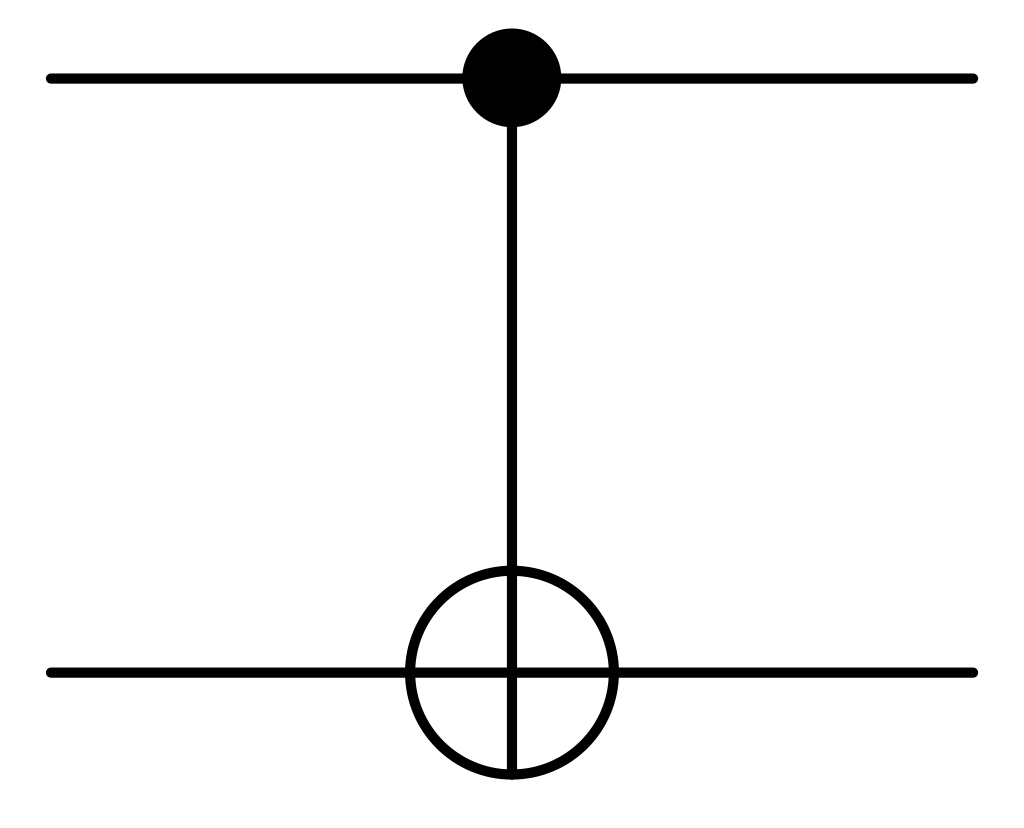}
  \caption{Controlled Pauli gate (CNOT)}
  %\label{fig:sub1}
\end{subfigure}%
\begin{subfigure} {0.5\linewidth}
  \centering
  \includegraphics[width=0.45\linewidth]{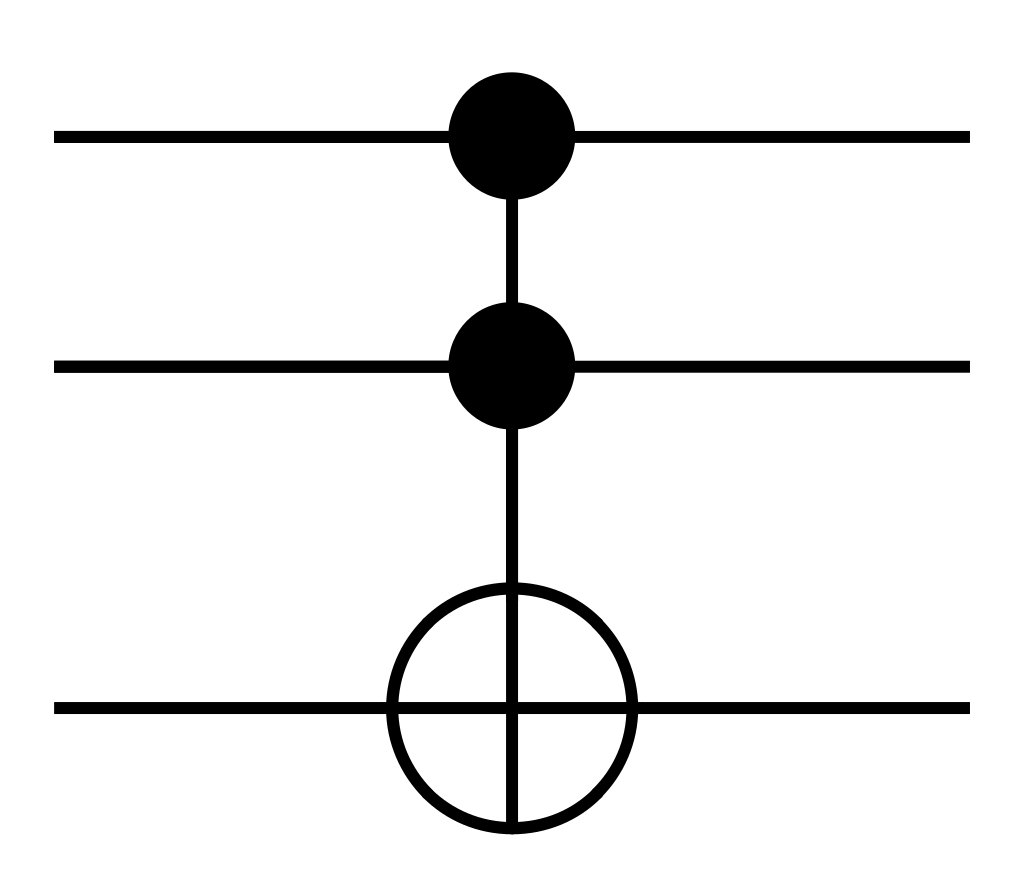}
  \caption{Circuit diagrams of Toffoli gate}
  \label{fig:sub2}
\end{subfigure}
\caption{Circuit representation of multi-qubit gates}
\label{fig:gates}
%\vspace{-0.5cm}
\end{figure}

In addition to 2-qubit gates like the CNOT gate, we also rely on 3-qubit controlled gates to implement logic locking functions. The Toffoli gate, also known as the Controlled-Controlled-NOT (CCNOT) gate, is a crucial multi-qubit gate in quantum computing. It extends the concept of the CNOT gate by involving three qubits: two control qubits and one target qubit. The Toffoli gate flips the state of the target qubit only when both control qubits are in the $\ket 1$. If either or both of the control qubits are in the $\ket 0$ state, the target qubit remains unchanged. 
% The significance of the Toffoli gate lies in its ability to perform reversible classical logic operations, making it a universal gate for classical reversible computing. It plays a crucial role in quantum algorithms that require conditional operations based on multiple qubits, such as error correction codes and arithmetic operations like multiplication. Its ability to execute multi-controlled operations makes the Toffoli gate an essential tool in constructing sophisticated logic locking systems.
3-qubit quantum gate enables us to use a controlled gate (e.g. the CNOT) as the target gate when locking quantum circuits.

\subsection{Related Work and Limitations}

Quantum circuits face significant security challenges, particularly with the rise of third-party quantum circuit compilers and quantum computer hardware in unsecured cloud environments. Potential attacks on quantum systems include, but are not limited to: insertion and alteration of quantum functions through trojans \cite{das2023trojannet,roy2024hardware}, crosstalk induced by fault injections~\cite{ash2020analysis}, side-channel information leakage by monitoring pulse power \cite{xu2023exploration, trochatos2023hardware}, and the threat of untrusted compilers leading to cloning, counterfeiting, and reverse engineering of quantum designs \cite{yang2024multi}. These security threats must be addressed at multiple levels, including hardware, compilation, and cloud infrastructure.

Quantum circuits are an important type of IP \cite{aboy2022mapping} and face IP threats similar to classical IC designs. The IP threats to quantum circuits in the compilation process are the major threats considered in this paper.
The compilation process ensures that every gate in the quantum circuit design is compatible and can be executed with the quantum computer. In addition, the compiler can apply optimizations to reduce errors and overhead of the quantum circuit. However, while compilation is necessary and beneficial, it inevitably exposes the quantum circuit to the compiler, posing a risk to the IP contained within the quantum circuit.
To protect the IP in the circuit, before sending it for compilation, the designer can obfuscate the circuit by transforming the circuit to a form from which the compiler cannot infer the original circuit. After compilation, the obfuscation is reversed on the compiled version to obtain a deobfuscated circuit, which must be functionally equivalent to the original design.

Existing work in quantum circuit obfuscation can be divided into three categories: (a) inserting random reversible gates into the quantum circuit \cite{das2023randomized, suresh2021short}, (b) splitting the quantum circuit into two or more sub-circuits and compiling separately  \cite{saki2021split}, and (c) adding extra key qubits that control quantum gates in the original circuit \cite{topaloglu2023quantum}. In \cite{das2023randomized}, a random reversible quantum circuit was generated and inserted into the front, middle, or end of the original quantum circuit as a means of obfuscation. After compilation, the reverse circuit of the random circuit is inserted into the same location to recover the correct functionality. \cite{suresh2021short} used a similar strategy. The main drawback of type of technique is that the topology of the original circuit is still fully exposed, and all that the adversary needs to find out is the boundary between the original and inserted portions. 
Split compilation ensures no individual compilers can see the whole original quantum circuit, but can be vulnerable to collusion among compilers \cite{saki2021split}.
The technique introduced in \cite{topaloglu2023quantum} bears more similarity with classical logic locking techniques where the functionality of the circuit is controlled by multiple key bits. For each key bit, a qubit is added to the quantum circuit. Due to the high costs of qubits and the limits on the number of qubits in current quantum computers, it is impractical to secure a quantum circuit using this technique in real-world scenarios. In this work, we present \eloq\ that addresses both the functional and structural vulnerabilities of current quantum circuit obfuscation techniques.

\section{Threat Models}
\begin{figure}[htb]
\captionsetup{font=small} 
    \centering
    \includegraphics[width=0.5\textwidth]{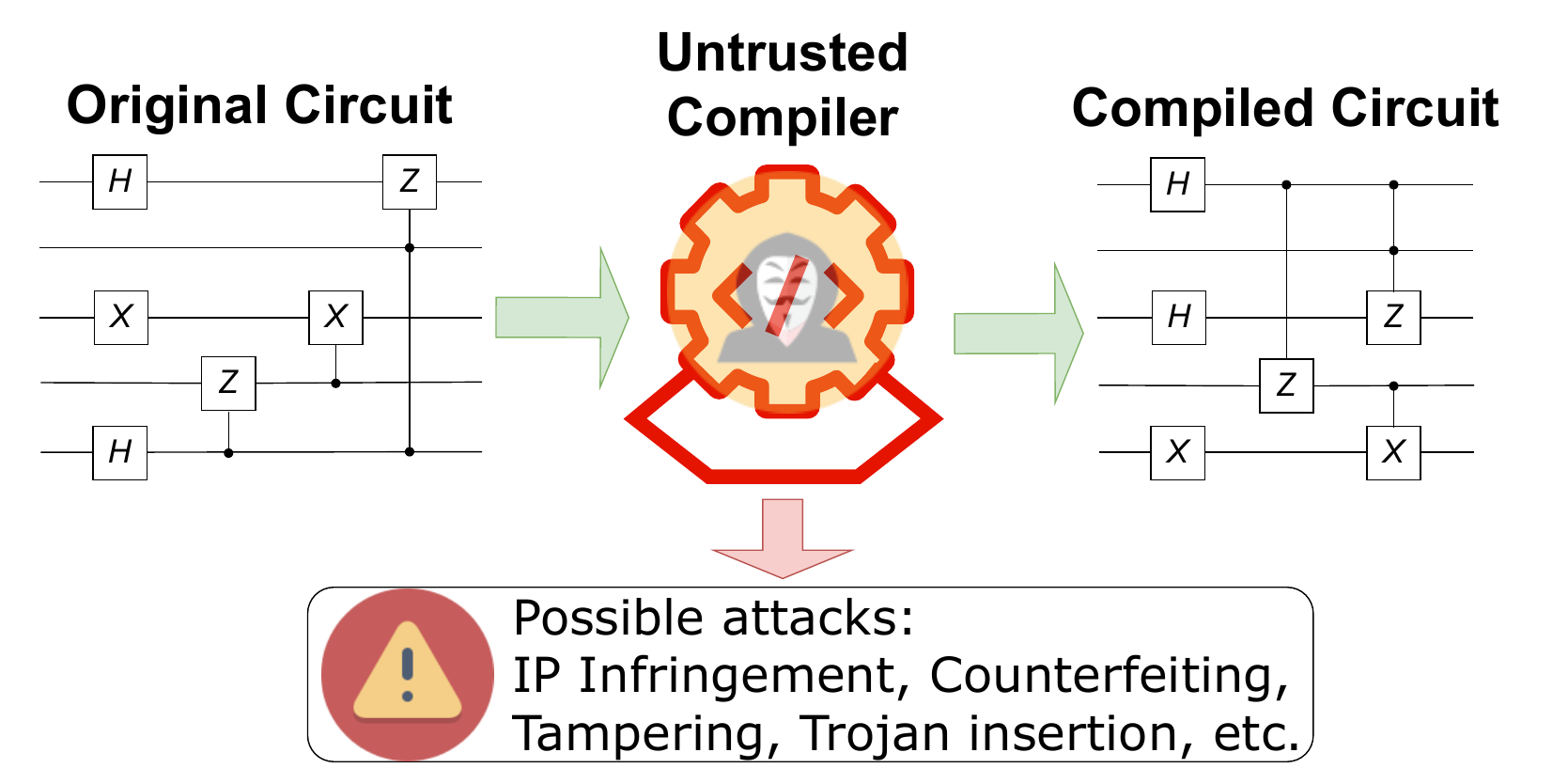}
    \caption{The untrusted compiler threat model. During the compilation process, the IP of the circuit is exposed and at the risk of IP theft. The circuit functionality may also be maliciously altered.}
    \label{fig:threat_model}
\end{figure}

% In this paper, we consider the third-party quantum compilers as potential adversaries. These third-party quantum compilers offer numerous benefits, such as support for multiple quantum computing platforms, advanced optimization, and error mitigation techniques, etc. \cite{smith2020open,salm2021automating}. Popular third-party quantum compilers include Qulic \cite{smith2020open}, TKET \cite{sivarajah2020t}, etc., and some hardware-specific compilers, like IBM's Qiskit \cite{qiskit2024} and Google's Cirq \cite{hancockcirq}, also support other platforms that can function as a third-party compiler. \footnote{These compilers are only listed as examples to demonstrate the popularity of third-party quantum compilers. We by no means suggest that these compilers are malicious.} Although third-party compilers offer some benefits, they also pose serious threats to the IP in quantum circuits. During the compilation process, the quantum circuit is fully exposed to the compiler, putting its IP at risk of being infringed or counterfeited. The compiler can also tamper with the design and make unauthorized changes that alter the functionality of the circuit. These threat models are illustrated in Figure \ref{fig:threat_model}. The threat model considered in this paper is consistent with those used in previous quantum adversaries approaches \cite{das2023randomized, suresh2021short, topaloglu2023quantum}. 
In this paper, we discuss the role of third-party quantum compilers as potential adversaries. Such compilers provide multiple benefits, including optimizing quantum circuits for a range of quantum computing platforms and applying error correction techniques \cite{smith2020open,salm2021automating}. Well-known examples of these compilers include Qulic \cite{smith2020open} and TKET \cite{sivarajah2020t}. Additionally, some compilers developed initially for specific quantum hardware platforms, such as IBM's Qiskit \cite{qiskit2024} and Google's Cirq \cite{hancockcirq}, also offer compatibility with other platforms, thus acting as third-party compilers. These compilers are mentioned merely to illustrate the prevalence of third-party quantum compilers. We are not implying they have malicious intents. Despite their advantages, third-party compilers introduce significant risks to the intellectual property (IP) of quantum circuits. The compilation process requires full disclosure of the quantum circuit to the compiler, exposing its IP to the possibility of counterfeiting. Compilers also have the potential to interfere with the design by making unauthorized modifications that could change the circuit's function. These potential threats are depicted in Figure \ref{fig:threat_model}. The threat model outlined in this paper aligns with those employed in previous studies of quantum adversaries \cite{das2023randomized, suresh2021short, saki2021split, topaloglu2023quantum}.

\begin{figure*}[htp]
\captionsetup{font=small} 
    \centering
      \includegraphics[width=0.7\textwidth]{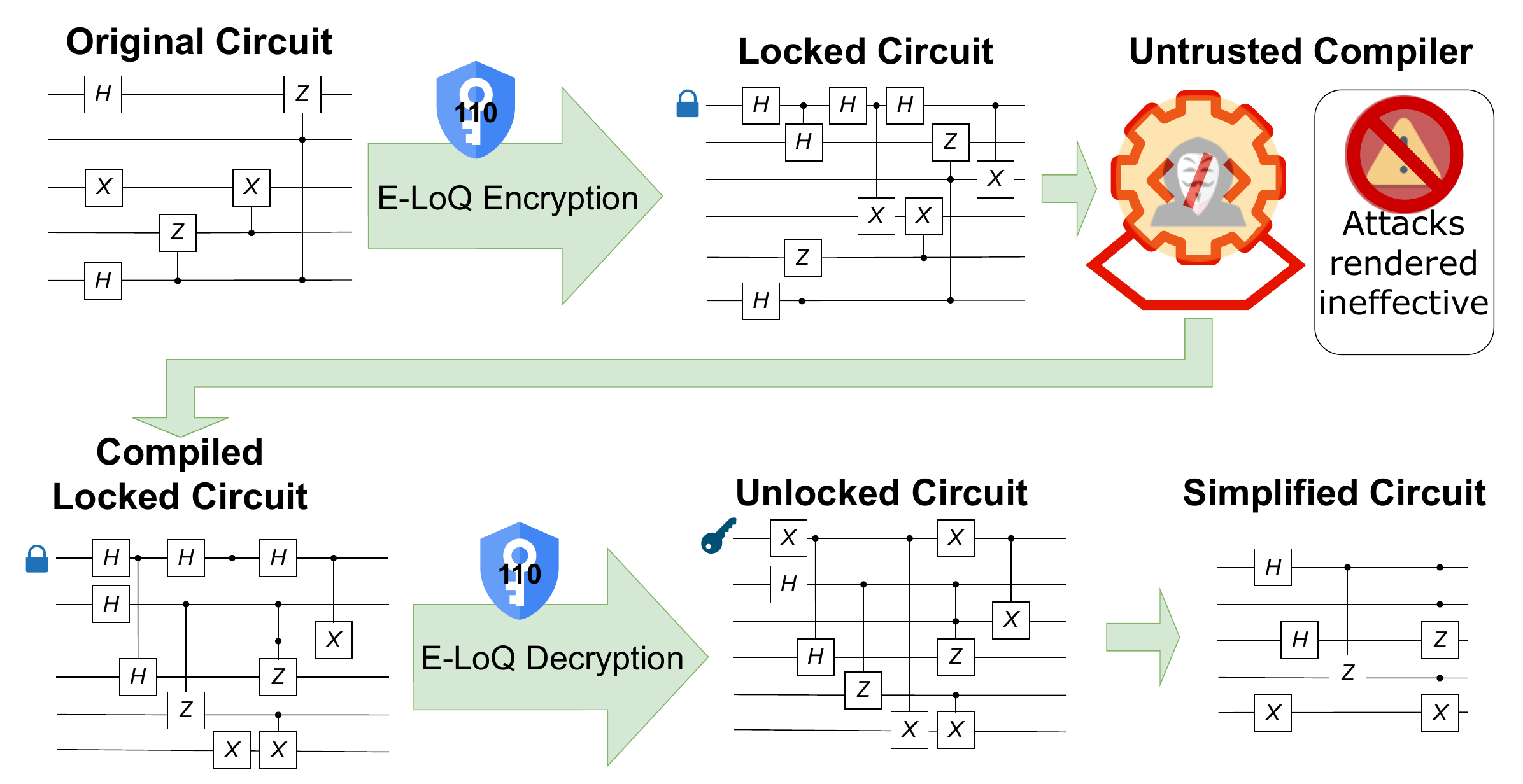}
    \caption{Overall defense flow using \eloq. Under this framework, the untrusted compiler only has access to the locked circuit which does not expose the key value. IP exposure is minimal and tampering is ineffective with unknown keys. The circuit can be decrypted and simplified post-compilation with the correct key, restoring functionality and reducing overhead.}
    \label{fig:defense_flow}
\end{figure*}

% During the compilation process, the correct key value is withheld from the compiler. To obtain the IP of the quantum circuit, the attacker needs to infer the original circuit $C$ or $\vec{k}$ from the locked circuit $C'$.

\section{Logic Locking for Quantum Circuits}

In general, our defense strategy focuses on limiting IP exposure through enhanced locking of quantum circuits (\eloq). The circuit is locked using a secret key similar to an encryption process. The ``encrypted'' circuit is then sent to compile. Since the untrusted compiler does not know the secret key, the correct functionality of the quantum circuit remains hidden from the compiler. After compilation, a ``decryption" process is applied to the encrypted circuit, restoring the functionality with the correct key. The unlocked circuit can then be simplified to reduce the overheads associated with locking. This flow is shown in Figure \ref{fig:defense_flow}. 

% to incorporate a locking key $\vec{k}$ and obtain obfuscated quantum circuit $C'_{\vec{k}}$. $C'_{\vec{k}}$ will only exhibit identical functionality to $C$ when the correct key is applied.

\subsection{Proposed \eloq\ Methodology}
\label{ssec:qll_method}
Suppose the original circuit is represented as $Circ$, with gates $G_0$, $G_1$, ..., $G_m$, where $m$ is the total number of gates. The circuit can be expressed as  $Circ=G_mG_{m-1}\cdots G_1G_0$ when considering the gates and the circuit as operators. Similarly, the locked circuit, represented as   $Circ'=G'_{m'}G'_{m'-1}\cdots G'_1G'_0$ where $G'_i$ are the gates in the locked circuit and $m'$ is the total number of gates.

\begin{figure*}[htb]
    \centering
    \begin{subfigure}{.25\linewidth}
        \includegraphics[width=0.90\linewidth]{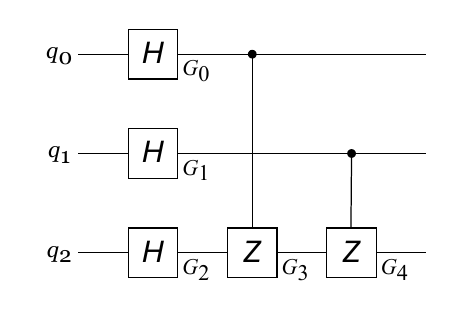}
        \caption{The original quantum circuit}
    \end{subfigure}
    \begin{subfigure}{.5\linewidth}
        \includegraphics[width=0.90\linewidth]{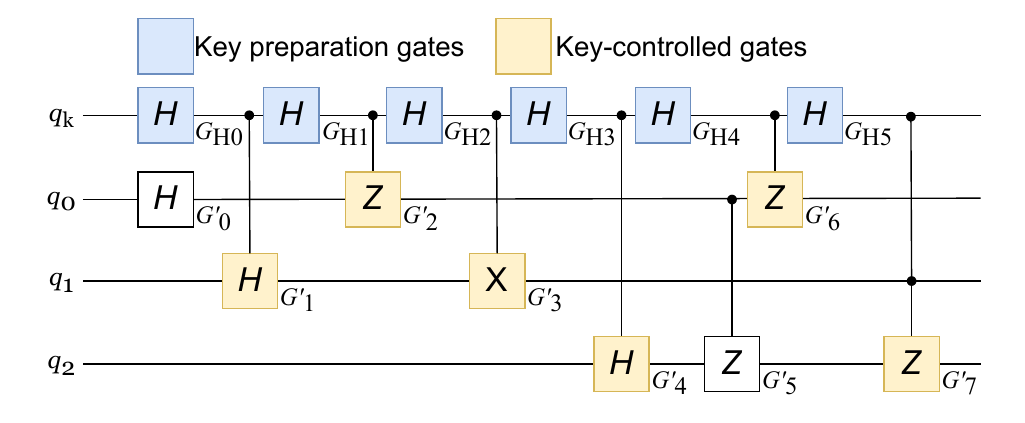}
        \caption{The locked quantum circuit}
    \end{subfigure}
    \begin{subfigure}{0.75\linewidth}
        \includegraphics[width=0.90\linewidth]{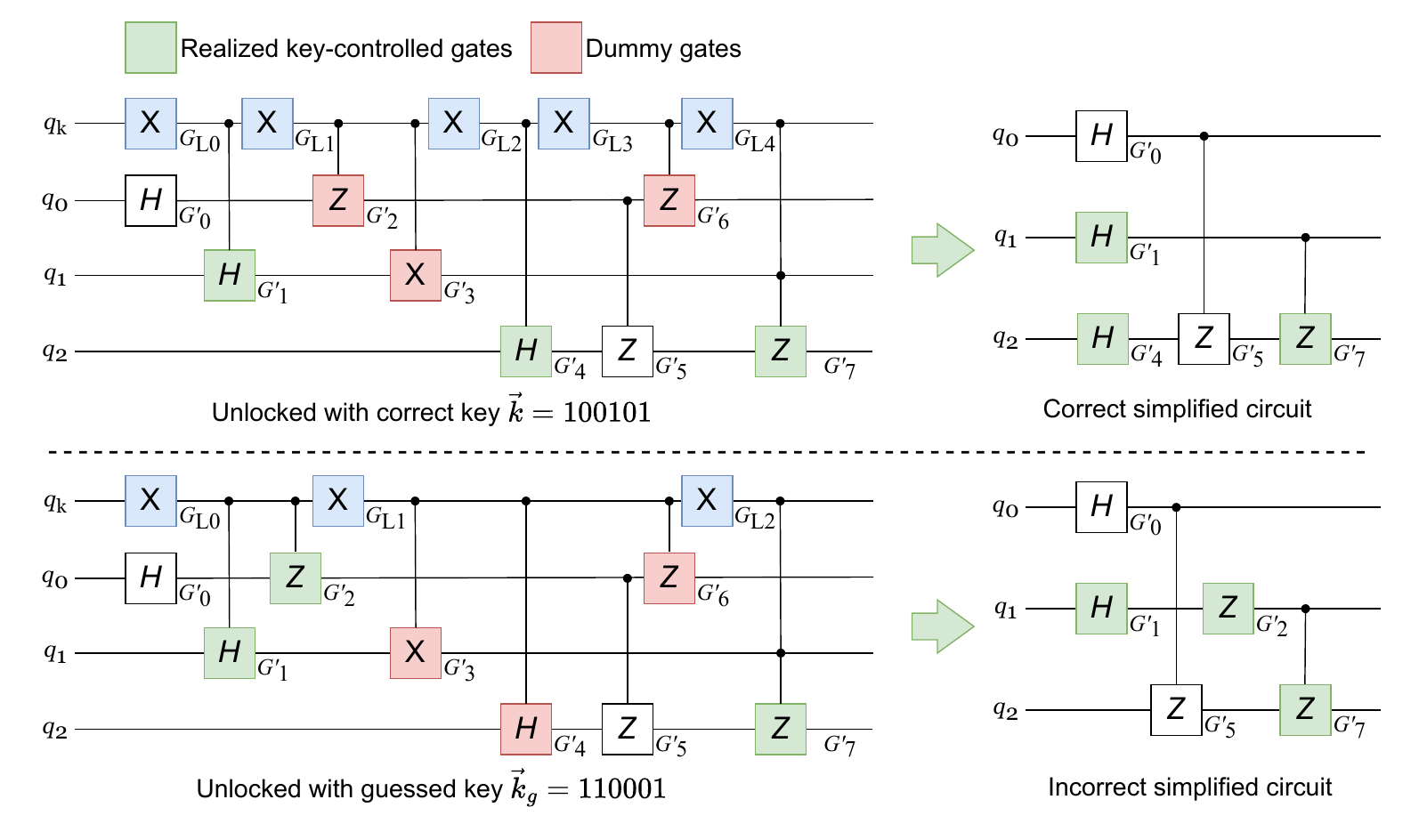}
        \caption{Compiled quantum circuit unlocked with correct key 100101 (upper left), guessed key 110001 (lower left), and their simplied circuits (right)}
    \end{subfigure}
    
    \caption{An example of \eloq. The original circuit in (a) is locked with a 6-bit key 100101 to get the locked citcuit (b). The locked circuit is then compiled and unlocked with two different keys: the correct key and a guessed key. Each unlocked circuit is then simplified. It can be seen in (c) that the two simplified circuits are different, indicating that not having the correct will make the adversary unable to know the functionality of the origianl circuit.}
    %\vspace{-0.6cm}
    \label{fig:qll_eg}
\end{figure*}

Our proposed \eloq\ technique includes three algorithms: $(\Gen,\Enc,\Dec)$. First, $\vec{k}=\Gen(1^n)$ selects an $n$-bit key value for $\vec{k}$ uniformly at random from $\{0,1\}^n$. Then, $Circ'=\Enc(Circ,\vec{k})$ transforms the original quantum circuit into the locked circuit. After compilation, $Circ''=\Dec(Circ',\vec{k})$ recovers the quantum circuit $Circ''$ from the locked circuit $Circ'$. To ensure the correctness of \eloq, we require $Circ= Circ'' =\Dec(\Enc(Circ,\vec{k}),\vec{k})$ for any circuit $Circ$ and the correct key $k$. In Figure \ref{fig:qll_eg}(a), a quantum circuit with 5 quantum gates ($m=5$) is shown to demonstrate the \eloq\ process. The detailed algorithms of \Enc\ and \Dec\ are described as follows with the example.

\subsubsection{\eloq\ Encryption Process}
$\Enc(Circ,\vec{k})$ begins by adding a key qubit $q_k$ into the quantum circuit. all key bits in $\vec{k}$ are implemented in this single qubit $q_k$, which is a significant improvement over previous approaches that use a separate qubit for each bit of key. Quantum gates are added on $q_k$ to facilitate the progression of key bit positions. When sending the encrypted circuit to the quantum compiler,  $q_k$ should not carry the correct key.
% A value of $|0\rangle$ indicates a 0 bit in $\vec{k}$ and $|1\rangle$ indicates 1. 
To maximally obfuscate the key, decorrelate adjacent key bits, and prevent the key-controlled gates from being simplified by the compiler, a Hadamard gate (H) is placed on $q_k$ before each quantum gate that is controlled by the correct key. We call this technique ``H-masking'' and these H gates will be removed and partially replaced with Pauli-X (i.e., NOT) gates to represent the correct key in the unlocking process. %Pauli-X (i.e., NOT) gates are added at appropriate locations on $q_k$ to represent the key value.
 
Modifications to the original circuit involve introducing two kinds of controlled gates into the circuit, both by $q_k$:
\begin{itemize}
    \item The first modification is to convert a subset of existing quantum gates into controlled gates, colored in green in Figure \ref{fig:qll_eg} (c). With the correct key, these gates should be controlled by $|1\rangle$, ensuring they function as the original circuit. 
    \item The second modification is to insert new controlled dummy gates that are governed by $|0\rangle$, colored in red in Figure \ref{fig:qll_eg} (c). Those gates behave as identity gates, leaving the functionality unchanged when the correct key is applied.
\end{itemize}

The configuration process works as follows. First, The number of 1's in $\vec{k}$ is calculated, i.e., $n_1=\sum_{i=0}^{n-1}\vec{k_i}$. Then $n_1$ out of the $m$ quantum gates in the circuit are randomly selected to be converted into controlled gates. \footnote{If $n_1>m$, then \Gen\ must be run again.} These gates will be controlled by $|1\rangle$ when the correct key is applied. The number of 0 is $n_0=n-n_1$, and for each 0 in the key, a new dummy controlled gate is inserted. The locations of these dummy gates are arranged in line with the order of the original controlled gates to ensure consistency with the key $\vec{k}$. % The type of dummy gates can be controlled Hadamard ($H$), CNOT ($CX$), controlled Pauli-Z ($CZ$), etc.

Shown as in the example of Figure \ref{fig:qll_eg}(a), we want to lock the original circuit with 6-bit key using \eloq\ ($n=6$), and the randomly generated key value $\vec{k}$ is 100101, i.e. $n_1=3$ and $n_0=3$. Therefore, 3 quantum gates in the original circuit need to be converted to controlled gates and 3 dummy gates need to be added to the circuit.
Let us assume that  $G_1$, $G_2$, and $G_4$ are randomly chosen to be converted to controlled gates. They become $G'_1$, $G'_4$, and $G'_7$ in the locked circuit shown in Figure \ref{fig:qll_eg}(b). Additionally, two dummy gates between $G'_1$ and $G'_4$, and another between $G'_4$ and $G'_7$, are added reflect the correct the 0's in the correct key sequence.

\subsubsection{\eloq\ Decryption Process}
$\Dec(Circ',\vec{k})$ is the decryption (unlocking) process in \eloq. First, the Hadamard gates on $q_k$ are removed and Pauli-X gates are inserted at appropriate locations on $q_k$ to ensure the correct functionality after unlocking.  In our case, since the key $\vec{k}$ starts with a 1, a Pauli-X gate is needed to prepare $q_k$ in the state $\ket 1$. As the next key bit is 0, another Pauli-X gate toggles $q_k$'s state back to $\ket 0$. In general, a Pauli-X gate is inserted on $q_k$ whenever the next key bit differs from the current one. 

Using this process, the locked circuit in Figure \ref{fig:qll_eg}(b) can be unlocked. On the left side of Figure \ref{fig:qll_eg}(c), we show the resulting circuits of unlocking the locked circuit with two different keys: the correct key $\vec{k}$ and a guessed key $\vec{k}_g$. 
In these circuits, the gates in green are controlled by $\ket 1$ and will perform the functionality of their non-controlled counterpart. Those in red, on the other hand, are controlled by $\ket 0$ and will act as identity gates whose output is equal to the input.
The unlocked circuit can be further simplified because the behavior of the controlled gates is already determined by the key. Gates controlled by $\ket 1$ can be replaced with an independent gate, and those controlled by $\ket 0$ can be removed. The key qubit $q_k$ can also be removed as it is not part of the original circuit. The simplified version of each unlocked circuit is shown to its right, where only the quantum gates in white and green remain.

One can also observe that the upper simplified circuit, the one corresponding to the correct unlocking key, is indeed the same as the one in Figure \ref{fig:qll_eg}(a), indicating that \eloq\ retains the circuit's functionality when the same key is used for encryption and decryption and causes minimal overhead to the compiled circuit. Without knowing the correct key, the attacker may attempt to unlock the circuit with a key guess $\vec{k}_g$. In this case, the resulting unlocked circuit will be $Circ''=\Dec(\Enc(Circ,\vec{k}),\vec{k}_g)$. For example, in the lower row of Figure \ref{fig:qll_eg}(c), we show a circuit unlocked with $\vec{k}_g=110001$ and its simplified version to its right. The simplified circuit is different from the original circuit. Hence, without the correct key, the attacker cannot unlock the compiled circuit correctly, rendering the attacks ineffective. 

\section{Metrics}
We propose two categories of evaluation metrics: quality metrics and security metrics. The quality metrics measure the functional corruption introduced by the locking mechanism when an incorrect key is used, whereas the security metric measures the difficulty of the attacker in finding the correct key or original circuit.

\subsection{Metrics for Evaluation Locking Quality}

\subsubsection{Total Variation Distance (TVD)}
TVD is a measure of the distance between two probability distributions in probability theory. This is highly relevant to quantum computing because unlike traditional computers, which produce deterministic results, the output of a quantum computing circuit is inherently probabilistic. For example, the output of a 1-bit circuit from a simulation, accounting for noise, can be represented as a distribution, such as \{``0'': 95, ``1'': 5\}, indicating the results of 100 shots. 
TVD can measure the difference in output distributions between the correct and locked circuits, thus indicating the effects of locking. We adopt the TVD formula, which was originally designed to work with continuous probability density functions to discrete distributions such as those of quantum circuit outputs. Specifically, TVD is calculated as the sum of the absolute differences between the counts of each outcome in the output distributions of the obfuscated and original circuits, normalized by the total number of shots. TVD can be expressed as follows: 
\begin{equation}
    TVD = \frac{\sum_{i=0}^{2^b-1} |y_{i,lock} - y_{i,orig}|}{2N}
\end{equation}
Where $N$ represents the total number of shots in this run, $b$ is the number of bits in the output, $y_{i,lock}$ and $y_{i,orig}$ represent the total number of measurement outcomes of value $i$ in the locked and original quantum circuits, respectively.

\subsubsection{Hamming Variation Distance (HVD)}
TVD metric measures the overall difference in counts. To compare the multi-bit output of the circuits, the counting of bits that are different, a.k.a. \textit{Hamming Distance (HD)}, is also very important. HD is not fully captured in TVD. For example, if the original circuit's output is 00 for a given shot, the contribution to TVD will be the same if the locked circuit's output is 01, 10, or 11. However, the output value of 11 has a higher HD than the original circuit's output, indicating a greater difference in circuit functionality.
Hence, if we only measure the count difference between the correct and obfuscated circuit, we overlook the valuable information in the Hamming distance between these outputs. To address this, we propose incorporating the Hamming distance into the TVD calculation and defining the Hamming Variation Distance (HVD). 
%When the output is multi-bit, the resulting distribution is a tuple such as 2-bit case \{``00'': 90, ``01'': 10, ``10'': 5, ``11'': 5\}. In such cases, the previously mentioned Total Variation Distance (TVD) only measures the absolute difference in counts without considering the Hamming distance between the output. For instance, in a 1-bit adder, both the carry and sum bits are measured, producing outputs like \{00, 01, 10, 11\}. If we only measure the absolute count difference between the correct and obfuscated circuit, we overlook the important information provided by the Hamming distance between these outputs. To address this, we propose incorporating the Hamming distance into the TVD calculation and defining the Hamming Variation Distance (HVD). 
Essentially, to calculate HVD, we apply the TVD formula on each output bit to calculate each output bit's contribution to the total HD. The formula is as follows:
% \begin{equation}
%     HVD = \frac{\sum_{i=0}^{2^b-1} \mathbf{HM}(x_{i,lock} - x_{i,orig}) * |y_{i,lock} - y_{i,orig}|}{2N}
% \end{equation}
\begin{equation}
    HVD = \frac{\sum_{j=0}^{b-1}\sum_{i=0}^{1} |y_{ji,lock} - y_{ji,orig}|}{2N}
\end{equation}
The definitions of $N$ and $b$ are the same as those for TVD. $y_{ji,lock}$ and $y_{ji,orig}$ represent the total number of measurement outcomes where the value of bit $j$ equals $i$ in the locked and original quantum circuits, respectively. 
% Specifically, the Hamming distance between two outputs $x_{lock}$ and $ x_{orig}$ (denoted as $\mathbf{HM}$) is multiplied by the absolute difference in counts ($|y_{lock}-y_{orig}|$) between the original and obfuscated circuits. 
This modification captures the significance of bit flips, meaning that the more bits that differ between the outputs, the more substantial the difference will be reflected in this metric. This enhanced measure provides a more comprehensive assessment of the impact of obfuscation on multi-bit quantum circuits.

\subsubsection{Degree of Functional Corruption (DFC)}

The Degree of Functional Corruption (DFC) is a metric introduced by \cite{das2023randomized} to quantify the extent to which a circuit's output has been obfuscated, compared to its correct, noise-free output. DFC measures the difference between the correct output and the highest incorrect output after obfuscation, with the results normalized by the total number of measurement shots, producing a value that ranges between -1 and 1. 

\begin{equation}
    DFC = \frac{Count_{|correct} - Count_{|incorrect}}{N}
\end{equation}

In an unobfuscated circuit, most measurement shots will yield the correct output, resulting in a DFC value close to 1. However, when a circuit is effectively obfuscated, the correct output bin will contain few or no counts, while the incorrect output bins will accumulate the majority of the counts. In such cases, the DFC value approaches -1, indicating high functional corruption. Therefore, a lower DFC value (closer to -1) signifies more effective obfuscation, as it indicates that the original circuit's functionality has been successfully concealed.

\begin{figure*}[!htp]
    \centering
    \captionsetup{font=small} 
    \includegraphics[width=0.95\textwidth]{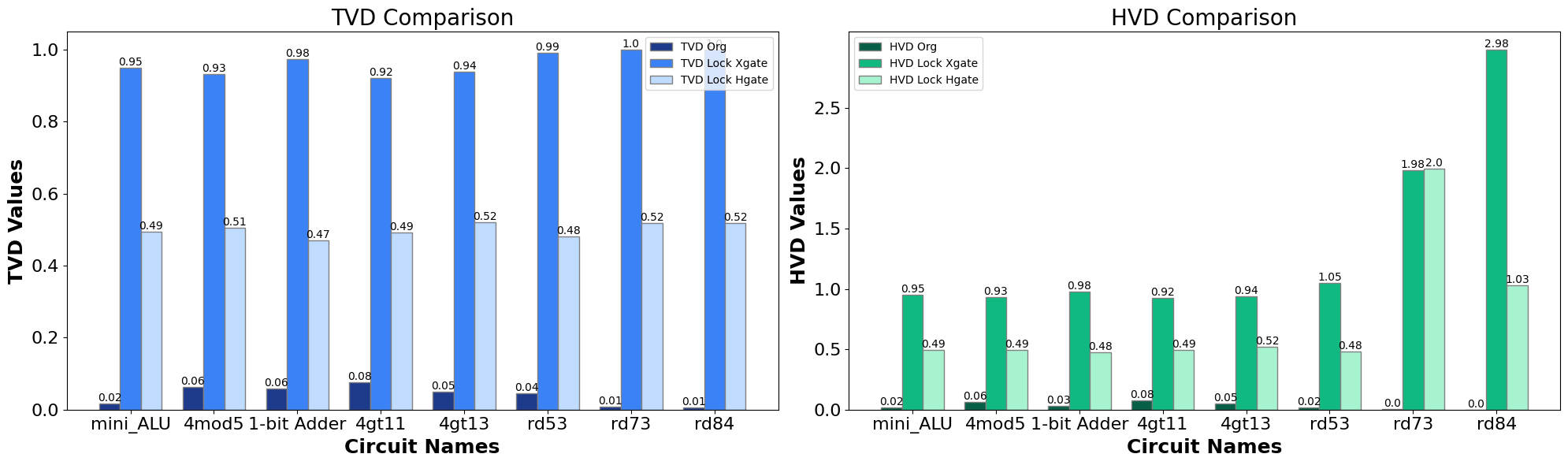}
    \caption{Distribution of Variation Distance (VD) of benchmark circuits: Both Total VD and Hamming VD are calculated and shown respectively. Selected circuits are simulated using Qiskit and FakeValencia backend, incorporating noise into the simulation}
    \label{fig:result_vd}
\end{figure*}

\subsection{Security Metrics}
% The security of \eloq\ can be characterized from two angles: functional corruption and structural obfuscation. Prior work focused primarily on functional corruption as the security metric for quantum circuit obfuscation. For example, two metrics were proposed in \cite{das2023randomized}. The \textit{total variation distance (TVD)} metric quantifies the output difference between the original and locked circuits. The \textit{degree of functional corruption (DFC)} is similar to TVD but compares the locked circuit with an error-free circuit. These metrics have several drawbacks. First, these metrics do not account for different key values an attacker may choose. An attacker may try different key values to unlock the circuit, and the locking mechanism should have a high error impact for any incorrect key. 
The security metrics for classical logic locking techniques are generally related to the resiliency to known attacks that are discussed in Section \ref{ssec:logic_locking}. However, none of the attacks on classical logic locking would apply to \eloq. 
Here, we propose a scheme to evaluate how effectively a quantum circuit is locked by key values. % The evaluation process is illustrated in the figure. 
Suppose we have a quantum circuit with an additional qubit key input. This additional key-qubit provides multiple positions where key values can be inserted. The $x$ is defined as the input of the circuit, and the output $y$ is measured as usual at the end of the circuit. In our evaluation, we first select different key inputs $k$, then traverse all possible inputs $x$, comparing the outputs under the same conditions with the correct key value, denoted as $k_{correct}$. This comparison gives us the guessing rate for each input. Finally, we average the guessing rates across all input sets to determine the overall guessing rate.

\begin{equation}
     guessRate = \frac{\sum_k ^ {N_k} \sum_{i} ^ {N_x} |y_{k,x_i} - y_{k_{correct},x_i}|} {N * N_k *N_x}
\end{equation}

\section{Evaluation}
\subsection{Experimental Setup}

We conducted our simulations using the IBM Qiskit framework to compile and simulate quantum circuits. The circuits were constructed using the \textit{QuantumRegister} and \textit{QuantumCircuit} modules. For our experiments, we utilized benchmark circuits from the \textit{RevLib} benchmarks \cite{wille2008revlib}, which has been widely used in prior work on quantum circuit compilation. These benchmark circuits encompass a variety of gate operations, with the number of gates ranging from 5 to 50 and qubit sizes varying across 4, 5, 7, 10, and 12 qubits. The benchmark circuits have
both 1-qubit output and multi-qubit outputs to simulate the function locking with different output cases.

To ensure realistic simulation conditions, we employed the FakeValencia backend from Qiskit \cite{qiskit2024}, which incorporates the noise model of the actual \textit{ibmq-valencia} device. All simulations were performed with 1,000 shots to generate statistically significant results. Both the original and the locked circuits were simulated using the same backend, ensuring that any differences observed are attributable to the locking mechanism rather than variations in the simulation environment.

As described in Section \ref{ssec:qll_method}, a dummy gate is inserted for each key bit of value 0. We strategically selected gate types for dummy insertion based on the operations present in the benchmark suite. For instance, in the \textit{RevLib} benchmarks, which predominantly feature arithmetic operations such as adders, ALUs, counters, and comparators, we primarily use CNOT gates to lock the circuit and obscure the correct input from potential attackers. For other types of circuits, such as those implementing Grover's algorithm, we opted to insert controlled Hadamard (H) gates. This tailored approach ensures that the locking mechanism is appropriately aligned with the nature of the operations in each circuit, reducing structural leakage and enhancing the effectiveness of the obfuscation.
% In summary, we evaluated two approaches for locking circuits: using logic gates (specifically X gate locking) and H gate locking. We will present our simulation results for both methods.

\subsection{Logic Locking Quality Analysis}
In this section, we present our simulation results, starting with the outcomes from the \textit{RevLib} benchmarks simulated using the Qiskit backend. These results encompass both 1-bit and multi-bit logic locking scenarios.

%\subsection{Fidelity Evaluation}

Figure \ref{fig:result_vd} illustrates a comparison of several metrics across different circuits, focusing on TVD (Total Variation Distance), HVD (Hamming Variation Distance). 
When presented with the locked circuit, the adversary potentially has two strategies when the key is unknown: to convert a subset of the H gates on $q_k$ into X gates to emulate a random circuit, or to keep the H hates as is and accept a 50\% error rate for each key bit. These two cases are referred to as ``Xgate'' and ``Hgate'' respectively for the two cases in Figure \ref{fig:result_vd}.
% In the left portion of Figure \ref{fig:result_vd}, we present the TVD values for three cases: the original circuit, the circuit the circuit locked using the X gate, and the circuit locked using the H gate. 
We will discuss the metrics obtained in the X gate case in this section, and the results for the H gate case will be illustrated in session \ref{sec:H masking}. TVD is calculated as the variation distance with the theoretical output, assuming no noise is present.  For instance, in the case of a 1-bit Adder with 100 shots, we use the outcome as  \{``0'': 100, ``1'': 0  \} as the reference to compare relative distance. The TVD values for most circuits are consistently close to 1, particularly in the benchmarks modified with the X gate, indicating a significant variation between the distributions. This consistency across circuits suggests that the alterations have a pronounced impact.

%However, in some circuits like 'mini\_ALU' and 4mod5, the original TVD values are lower, which might indicate less variation in those specific cases.

Similarly, the HVD values demonstrate a significant divergence in the altered benchmarks, with the original HVD values being relatively low in comparison.  This indicates that the alterations have markedly impacted the distributions measured by HVD. Notably, circuits with multiple outputs, such as the 1-bit Adder, rd53, rd73, and rd84, exhibit comparable higher HVD values. This is because HVD measures the bit flips from the original output, accumulating over the number of bits.  In these cases,  HVD value would exceed 1 as multiple-bit are computed together.  In the case of 1-bit outputs, the TVD and HVD are equal.

%In these cases, divergence is observed because the altered output partially overlaps with the correct output. For example, if the correct outcome is ``0000'', and the altered outcome is ``0001'', the HVD would only account for a 1-bit difference.
\begin{figure}[h]
    \centering
    \captionsetup{font=small} 
    \includegraphics[width=0.48\textwidth]{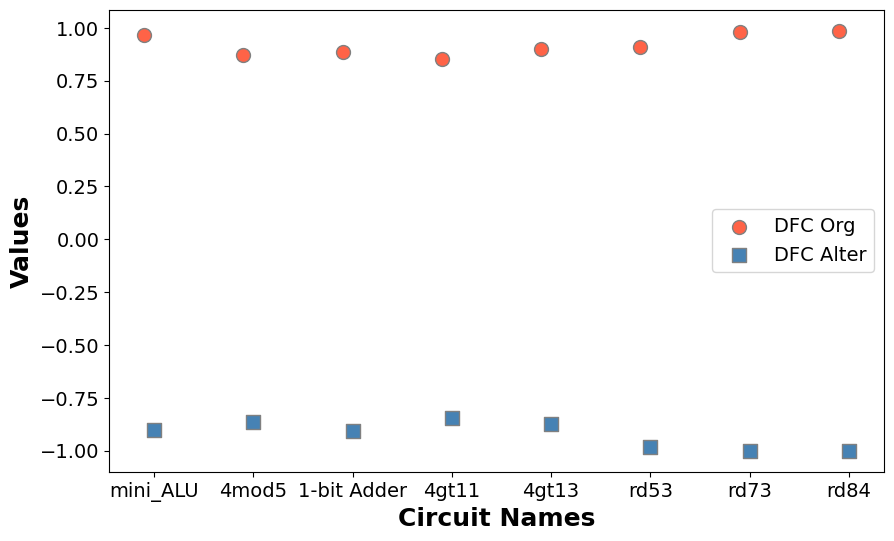}
    \caption{Degree of Functional Corruption (DFC) values for selected benchmark circuits \cite{wille2008revlib} are presented with a range from -1 to 1; Values near 1 indicate that the function remains as designed, while a value near -1 signifies a deviation from the intended function. The original unlocked circuit precisely presents the designed function, whereas the altered circuits demonstrate varying degrees of function corruption. }
    \label{fig:result_dfc}
\end{figure}

Overall, the figure shows that the altered benchmarks diverge significantly from the original benchmarks across all metrics. The consistent and high values in the altered benchmarks indicate that the changes had a uniform impact across different circuits, affecting all of them in a similar way. 

%This highlights the systematic nature of the alterations and their substantial effect on the data distributions across various circuit types.

Degree of Functional Corruption (DFC) values for selected benchmark circuits \cite{wille2008revlib} are presented with a range from -1 to 1; Values near 1 indicate that the function remains as designed, while a value near -1 signifies a deviation from the intended function. Figure \ref{fig:result_dfc} presents the Degree of Functional Corruption (DFC) values across different circuits, comparing three benchmarks: DFC Org (Original) and DFC Alter (Altered). The original DFC values are consistently high across all circuits, close to or at 1.0, indicating that the original simulation results closely matched the theoretical outputs. In contrast, the altered DFC values are markedly different, with all values being negative and close to -1, suggesting that the alterations had a significant impact on the circuit outcomes.

%The original unlocked circuit precisely presents the designed function, whereas the altered circuits demonstrate varying degrees of function corruption.

%\subsection{Fidelity Evaluation}

%The Degree of Functional Corruption (DFC) is a metric introduced by \cite{das2023randomized} to quantify the extent to which a circuit's output has been obfuscated, compared to its correct, noise-free output. DFC measures the difference between the correct output and the highest incorrect output after obfuscation, with the results normalized by the total number of measurement shots, producing a value that ranges between -1 and 1. 

%The Multi-bit DFC values, while positive, are significantly lower than those of the original DFC. This implies that the locking circuits partially obscured the functionality of the multi-bit outputs. It indicates that fully masking the functionality of multi-bit output circuits (e.g., 1-bit adders) is more challenging than doing so for single-bit output circuits, likely requiring more insertions and alterations of the original circuits.

\subsection{Locking Results with H Masking} 
\label{sec:H masking}
When we use the X-gate or similar logic quantum gates for locking circuits, it effectively functions as a bit-flip operation on the original output, inverting the bit with the insertion of a controlled bit. However, this bit-flipped output remains highly correlated with the original input, with a correlation coefficient of approximately -1. To neutralize this strong correlation, we replace the X-gate with the H-gate for locking. The H-gate places the qubit in an equal probability of being measured as 0 or 1, reducing the direct correlation with the original output of the circuit. 
By using the H-gate for locking, we observed similar metrics, such as TVD and HVD values. In the TVD subfigure of Figure \ref{fig:result_vd}, the original TVD values are consistently low across all circuits, indicating minimal variation. However, after applying the H-gate locks, the TVD values increase significantly, though not as much as with the X-gate, suggesting a more balanced (50:50) impact on the distribution by the H-gate lock. 
Similarly, in the HVD subfigure, the results after applying the locks with H-gate lead to a more moderate increase, balancing the distribution around 0.5 rather than fully inverting it. Overall, H-gate locks increase both variation and divergence, as indicated by the rise in TVD and HVD values, creating a neutralized and unpredictable outcome that makes it more challenging for an adversary to deduce the correct circuit function.

\subsection{Cost and Overhead Analysis}

\subsubsection{Gate Counts and Circuit Depth}

The comparison of metrics across different circuits, as shown in Table \ref{tab:circuit_parameters}, reveals several trends in how alterations affect circuit complexity and performance. Generally, the insertion of additional controlled keys increases circuit complexity, as evidenced by slight increases in both depth and gate count in most cases. However, for deeper and larger circuits, the impact of insertion is minimal, since the absolute number of additional gates is small relative to the original gate count. In some instances, inserting a controlled gate into an empty slot does not alter the circuit depth, which is particularly evident in circuits like rd53 and rd84, where the depth remains unchanged after the alteration.

\subsubsection{Fidelity of Unlocked Circuits}

In this section, we evaluate the performance of circuits after being unlocked by the correct keys and simplified. Introducing additional gates and layers can affect a circuit's fidelity, particularly in the presence of noise. Therefore, we assessed the accuracy (the ratio of correct outcomes to the total number of shots) both before and after the insertion of quantum gates. As shown in Table \ref{tab:circuit_parameters}, most circuits experience only a slight decrease in accuracy after the alterations, with fidelity changes typically remaining under 1\%. The result indicates that circuits with fewer gates and shorter depths are more sensitive to these alterations, while larger circuits are less affected. Overall, although the alterations generally increase circuit complexity, the impact on performance is minimal, indicating that the circuits remain robust even after the changes are introduced.

\begin{table*}[h]
\centering
\begin{tabular}{|l|c|c|c|c|c|c|c|}
\hline
\textbf{Circuit} & \textbf{circuit depth} & \textbf{depth\_alter} & \textbf{Gate count} & \textbf{count\_alter} & \textbf{accuracy} & \textbf{accuracy\_after} & \textbf{Fidelity change} \\ \hline
mini\_ALU & 8  & 9  & 9  & 11 & 0.985 & 0.975 & 1.02\% \\ \hline
4mod5    & 5  & 6  & 6  & 8  & 0.94  & 0.93  & 1.06\% \\ \hline
1-bit adder & 5  & 5  & 7  & 9  & 0.967 & 0.951 & 1.65\% \\ \hline
4gt11    & 13 & 13 & 13 & 15 & 0.937 & 0.931 & 0.64\% \\ \hline
4gt13    & 4  & 6  & 4  & 6  & 0.95  & 0.941 & 0.95\% \\ \hline
rd53     & 16 & 16 & 19 & 21 & 0.945 & 0.94  & 0.53\% \\ \hline
rd73     & 13 & 15 & 23 & 26 & 0.993 & 0.986 & 0.70\% \\ \hline
rd84     & 15 & 15 & 32 & 34 & 0.996 & 0.993 & 0.30\% \\ \hline
\end{tabular}
\caption{Comparison of circuit parameters: depth, count, accuracy, and fidelity change before and after alterations. The original circuit is from the \textit{RevLib}.}
\label{tab:circuit_parameters}
\end{table*}

\subsection{Analysis of Key Guessing}

\begin{figure}[h]
    \captionsetup{font=small} 
    \centering
    \includegraphics[width=0.48\textwidth]{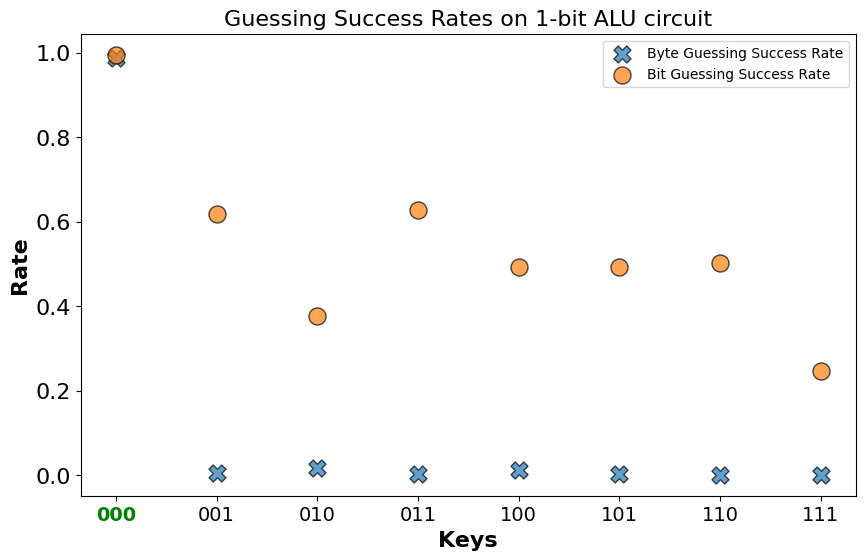}
    \caption{Key guessing rate on 1-bit ALU circuit; the X-axis displays the key values, with  ``000" as the correct key. Byte guessing rate, indicated by ``X" represents the rate of the correct output pattern compared to the maximum output from locked circuits; Bit rate is represented by ``circle" and the value is averaged over all bits. }
    \label{fig:result_guess}
\end{figure}
To evaluate the success rate of the key guessing procedure, we collected results by traversing all possible key inputs. For instance, in the case of a 1-bit ALU locked with a 3-bit key, only the correct key  ``000" would produce the exact correct function of the circuit. We aimed to assess the output difference when the correct key is used versus when an incorrect key is applied. To achieve this, we simulated the circuit across all input cases and compared the output bits. The ALU outputs are ``01", and we compared the matching rate between the correct key output and the incorrect key outputs across different input patterns. A perfect match results in a rate close to 1, while a mismatch counts as 0. This corresponds to the key guessing rate shown in Figure \ref{fig:result_guess}

Since the output is a tuple, individual bits may match even if the entire output does not. Therefore, we calculate the bit-to-bit match rate and average it across all bits, resulting in the bit-guessing rate. For example, in the case of the ALU, which has a two-bit output (Carry and Sum), we first compare the Carry bit between the correct and incorrect keys, then do the same for the Sum bit. This bit-by-bit comparison provides a more detailed assessment of the key guessing accuracy.

Figure \ref{fig:result_guess} shows the guessing success rates for both byte and bit-level guesses across different keys, represented by binary strings on the X-axis. The byte guessing success rate for the key ``000" is exceptionally high, close to 1, indicating a strong likelihood of correctly guessing this byte. In contrast, the other keys have much lower byte guessing success rates, suggesting a significantly reduced probability of correct guesses. The bit guessing success rates vary more across the keys, with ``000" still having the highest success rate. The rates for other keys are close to 0.5, resembling the randomness of a bit guess. This figure effectively illustrates the distribution of guessing success rates across different keys, demonstrating that our method can successfully lock the correct function of the circuit.

\section{Discussion and Future Work}
We evaluated the performance of \eloq\ using numerical metrics, which provided insight into how the insertion of key bits affects circuit behavior and fidelity. 
The evaluation result shows that \eloq\ will cause substantial errors in the locked circuit and is resilient to key guessing attacks.
Compared to existing quantum circuit obfuscation approaches, \eloq\ has the following additional advantages:
\begin{itemize}
    \item Lower structural information leakage. In a \eloq-locked circuit, controlled gates are inserted at multiple locations. Since each controlled gate can be either real or dummy, the unaltered portion from the original circuit becomes segmented, and the attacker does not know the connectivity between them due to the potentially dummy controlled gates. In comparison, existing approaches \cite{das2023randomized, topaloglu2023quantum} leave large portions of the original circuit undisturbed, causing high structural leakage.
    \item Lower overhead. In our approaches, the added key qubit $q_k$ and dummy gates are only temporary for the outsourced compilation process. After compilation, in the decryption phase, all of them can be removed, making the overhead very minimal. In comparison, existing approaches will keep the added circuitry \cite{das2023randomized} or key qubit \cite{topaloglu2023quantum} permanently in the unlocked circuit, causing significant overhead.
\end{itemize}
In our future work, we will develop a quantitative structural leakage metric for structural information leakage. We also plan to demonstrate the efficacy of \eloq\ in real quantum hardware to better understand the impact of \eloq\ on noise and performance.

\section{Conclusion}

% The quantum compilers are key steps to optimizing circuits and translating high-level gates into hardware-compatible operations, however, they can also pose risks when untrusted, potentially leading to the theft of valuable quantum designs. To address this, we proposed \eloq, a technique inspired by classical logic locking used to secure traditional integrated circuits. Our method involves inserting controlled gates into quantum circuits, controlling their function based on the key value not exposed to malicious attackers. We expanded on previous quantum circuit locking methods by introducing a more efficient multi-bit key per qubit approach that utilizes various quantum gates.
The quantum compilers are key steps to optimizing circuits and translating high-level gates into hardware-compatible operations, however, they can also pose risks when untrusted, potentially leading to the theft of valuable quantum designs. To address this, we proposed \eloq\ which expands on  classical-logic-locking-inspired prior work by introducing a more efficient multi-bit key per qubit approach that utilizes various quantum gates, controlling their function based on the key value not exposed to malicious attackers. 
Our experiments on benchmark quantum circuits showed that \eloq\ effectively conceals the original circuit's function, as evaluated using metrics such as TVD, HVD, and DFC. Additionally, we demonstrated that using an incorrect key by an attacker significantly disrupts the circuit's functionality. Importantly, our modifications had minimal performance impact, with an average fidelity degradation of less than 1\%. This work provides a practical solution for securing quantum circuits against reverse engineering and IP theft in the rapidly evolving field of quantum computing.
\vspace{1cm}

%\bibliographystyle{ACM-Reference-Format}

\balance

\bibliographystyle{IEEEtran.bst}
\bibliography{qref}
%\bibliography{reference}

\end{document}